\documentclass[pdftex,nomathfonts]{aipproc} 
\layoutstyle{6s}
\pdfoutput=1
\usepackage{graphicx}

\title{Bayesian segmentation of hyperspectral images} 

\author{Adel Mohammadpour$^\dagger$}
 {address = {Department of Statistics,
 Faculty of Mathematics \& Computer  Science \linebreak Amirkabir University
 of Technology, 424 Hafez Ave., 15914 Tehran, Iran },
 email={\tt adel@aut.ac.ir}
 }
 
\author{Olivier F\'eron}
 {
  address = {Laboratoire des Signaux et Syst\`emes, (CNRS-Sup\'elec-UPS)
 \linebreak  
  Sup\'elec, Plateau de Moulon, 91192 Gif-sur-Yvette, France},
  email={\tt mohammadpour,feron,djafari@lss.supelec.fr}
 }
\author{Ali Mohammad-Djafari}
 {
  address = {Laboratoire des Signaux et Syst\`emes, (CNRS-Sup\'elec-UPS)
 \linebreak  
  Sup\'elec, Plateau de Moulon, 91192 Gif-sur-Yvette, France},
  email={\tt mohammadpour,feron,djafari@lss.supelec.fr}
 }
\date{}

\def\bdoc{\begin{document}}             \def\edoc{\end{document}}
\def\babs{\begin{abstract}}             \def\eabs{\end{abstract}}
\def\bcc{\begin{center}}                \def\ecc{\end{center}}
\def\ben{\begin{enumerate}}             \def\een{\end{enumerate}}
\def\bit{\begin{itemize}}               \def\eit{\end{itemize}}
\def\beq{\begin{equation}}              \def\eeq{\end{equation}}
\def\btabu{\begin{tabular}}             \def\etabu{\end{tabular}}
\def\barr{\begin{array}}                \def\earr{\end{array}}
\def\beqn{\begin{eqnarray}}             \def\eeqn{\end{eqnarray}}
\def\beqnx{\begin{eqnarray*}}           \def\eeqnx{\end{eqnarray*}}

\def\ab{{\bm a}}\def\Ab{{\bm A}}\def\Ac{{\cal A}}
\def\bb{{\bm b}}\def\Bb{{\bm B}}\def\Bc{{\cal B}}
\def\cb{{\bm c}}\def\Cb{{\bm C}}\def\Cc{{\cal C}}
\def\db{{\bm d}}\def\Db{{\bm D}}\def\Dc{{\cal D}}
\def\eb{{\bm e}}\def\Eb{{\bm E}}\def\Ec{{\cal E}}
\def\fb{{\bm f}}\def\Fb{{\bm F}}\def\Fc{{\cal F}}
\def\gb{{\bm g}}\def\Gb{{\bm G}}\def\Gc{{\cal G}}
\def\hb{{\bm h}}\def\Hb{{\bm H}}\def\Hc{{\cal H}}
\def\ib{{\bm i}}\def\Ib{{\bm I}}\def\Ic{{\cal I}}
\def\jb{{\bm j}}\def\Jb{{\bm J}}\def\Jc{{\cal J}}
\def\kb{{\bm k}}\def\Kb{{\bm K}}\def\Kc{{\cal K}}
\def\lb{{\bm l}}\def\Lb{{\bm L}}\def\Lc{{\cal L}}
\def\mb{{\bm m}}\def\Mb{{\bm M}}\def\Mc{{\cal M}}
\def\nb{{\bm n}}\def\Nb{{\bm N}}\def\Nc{{\cal N}}
\def\ob{{\bm o}}\def\Ob{{\bm O}}\def\Oc{{\cal O}}
\def\pb{{\bm p}}\def\Pb{{\bm P}}\def\Pc{{\cal P}}
\def\qb{{\bm q}}\def\Qb{{\bm Q}}\def\Qc{{\cal Q}}
\def\rb{{\bm r}}\def\Rb{{\bm R}}\def\Rc{{\cal R}}
\def\sb{{\bm s}}\def\Sb{{\bm S}}\def\Sc{{\cal S}}
\def\tb{{\bm t}}\def\Tb{{\bm T}}\def\Tc{{\cal T}}
\def\ub{{\bm u}}\def\Ub{{\bm U}}\def\Uc{{\cal U}}
\def\vb{{\bm v}}\def\Vb{{\bm V}}\def\Vc{{\cal V}}
\def\wb{{\bm w}}\def\Wb{{\bm W}}\def\Wc{{\cal W}}
\def\xb{{\bm x}}\def\Xb{{\bm X}}\def\Xc{{\cal X}}
\def\yb{{\bm y}}\def\Yb{{\bm Y}}\def\Yc{{\cal Y}}
\def\zb{{\bm z}}\def\Zb{{\bm Z}}\def\Zc{{\cal Z}}

\def\alphab{\bm{\alpha}}
\def\betab{\bm{\beta}}
\def\gammab{\bm{\gamma}}            \def\Gammab{\bm{\Gamma}}
\def\deltab{\bm{\delta}}            \def\Deltab{\bm{\Delta}}
\def\epsilonb{\bm{\epsilon}}
\def\varepsilonb{\bm{\varepsilon}}
\def\etab{\bm{\eta}}
\def\thetab{\bm{\theta}}            \def\Thetab{\bm{\Theta}}
\def\varthetab{\bm{\vartheta}}
\def\iotab{\bm{\iota}}
\def\kappab{\bm{\kappa}}
\def\varkappab{\bm{\vrkappa}}
\def\lambdab{\bm{\lambda}}          \def\Lambdab{\bm{\Lambda}}
\def\mub{\bm{\mu}}
\def\nub{\bm{\nu}}
\def\xib{\bm{\xi}}                  \def\Xib{\bm{\Xi}}
\def\pib{\bm{\pi}}                  \def\Pib{\bm{\Pi}}
\def\varpib{\bm{\varpi}}
\def\taub{\bm{\tau}}
\def\phib{\bm{\phi}}
\def\varpib{\bm{\varpi}}
\def\rhob{\bm{\rho}}
\def\varrhob{\bm{\varrho}}
\def\sigmab{\bm{\sigma}}            \def\Sigmab{\bm{\Sigma}}
\def\varsigmab{\bm{\varsigma}}
\def\phib{\bm{\phi}}                \def\Phib{\bm{\Phi}}
\def\varphib{\bm{\varphi}}
\def\chib{\bm{\chi}}
\def\psib{\bm{\psi}}                \def\Psib{\bm{\Psi}}
\def\omegab{\bm{\omega}}            \def\Omegab{\bm{\Omega}}
\def\taub{\bm{\tau}}
\def\uepsilonb{\bm{\uepsilon}}      \def\Uepsilonb{\bm{\Uepsilon}}
\def\zetab{\bm{\zeta}}

\def\fbu{\underline{\fb}}
\def\gbu{\underline{\gb}}
\def\epsilonbu{\underline{\epsilonb}}
\def\thetabu{\underline{\thetab}}
\def\bm#1{\mbox{\boldmath $#1$}}

\begin{abstract}
In this paper we consider the problem 
of joint segmentation of hyperspectral images in the Bayesian
framework.    
The proposed approach is based on a Hidden Markov Modeling (HMM) of the images
with common segmentation, or equivalently with common hidden classification label variables which is modeled by a Potts Markov Random Field.
We introduce   an appropriate Markov Chain Monte Carlo (MCMC) algorithm to implement the method and show some simulation results.

\end{abstract}

\begin{document}
\maketitle

\section{Introduction}
The most significant recent advances in remote sensing has been the development of 
hyperspectral sensors and software to analyze the resulting image data.  Over the past decade hyperspectral image analysis has 
matured into one of the most powerful and fastest growing technologies in the field of 
remote sensing. 
The "hyper" in hyperspectral means "over" as in "too many" and refers to the large 
number of measured wavelength bands. Hyperspectral images are spectrally 
overdetermined, which means that they provide ample spectral information to identify 
and distinguish spectrally unique materials. Hyperspectral imagery provides the potential 
for more accurate and detailed information extraction than possible with any other type of 
remotely sensed data. However the huge amount of data in
hyperspectral images make its information exploitation difficult and image processing tools (classification, segmentation, comprising and coding) are needed to summarize the information included in these data.
\par
This paper will introduce a segmentation method for hyperspectral images.
Several unsupervised and supervised 
algorithms have been developed for segmentation of 
multispectral images. However, these algorithms fail to deliver 
high accuracies for classifying hyperspectral images \cite{Jensen95,Jia99,Landgrebe02,Shaw02,Shah03}. 
In this paper we consider the problem 
of joint segmentation of hyperspectral images in the Bayesian
framework.    

The proposed approach is based on a Hidden Markov Modeling (HMM) of the images
with common segmentation, or equivalently with common hidden classification label variables which is modeled by a Potts Markov Random Field.
We introduce   an appropriate Markov Chain Monte Carlo (MCMC) algorithm to implement the method and show some simulation results.
This approach has previously been  considered by \cite{Feron04a} for multispectral images. 

In that work,  the  pixels of the same region  in different images 
are assumed independent.
This independence assumption is a valid hypothesis 
for multispectral images. However in  hyperspectral images the pixel values in each channel are not independent. This work is then an extension
to that work by considering a Markov model for these pixels along each channel.

This paper is organized as follows: In the next two sections first  we introduce our method for segmentation of hyperspectral images
in the Bayesian framework. Then we propose an appropriate MCMC Gibbs sampling particularly designed for this segmentation task. Finally, in the last section we present some simulation results to show the performances of the proposed methods.

\section{MODELING FOR BAYESIAN SEGMENTATION}
\label{model}
Let $g_i(r)$  be the observed value of the pixel $r$,   $r\in \Zb^2$, in the spectral band $i$ of a hyperspectral image. We model the observations by
\beq
\label{eq3}
g_i(r)=f_i(r)+\varepsilon_i(r), \qquad i=1,\dots,M
\eeq
\noindent
where $f_i(r)$ is the unknown perfect value of $g_i(r)$ and $\varepsilon_i(r)$
is a noise.
    Note that if we consider images, the pixels $r$ belong to a finite lattice $\Sc$, and we will note $S$ the number of pixels of this lattice. In the following we also use the notations
\beq
\gb_i = \fb_i + \varepsilonb_i \qquad or \qquad \gbu = \fbu + \epsilonbu ,
\label{modeldebase}
\nonumber
\eeq
\noindent
where $\gb_i = \{ g_i(r), r \in  \Sc \}$ and $\gbu = \{ \gb_i,i=1,\dots,M \}$
and a similar definition for $\fbu$ and $\epsilonbu$.
We introduce a label variable $z(r)$ for the regions and consider the region labels as common feature between all images.  Thus the  hidden variable $\zb = \{z(r),r \in \Sc\}$  represent a common classification of the images for different
bands. 
The main result of this paper is estimation of joint segmentation label $\zb $. 

Assuming independent noises $\varepsilonb_i$ among the different observations we have
\beq
p(\gbu|\fbu,\varepsilonb_i)  = \prod_{i=1}^M p(\gb_i|\fb_i) = \prod_{i=1}^M p_{\varepsilonb_i}(\gb_i-\fb_i).
\nonumber
\eeq
\noindent
Assuming $\varepsilonb_i $ centered, white and Gaussian $\varepsilonb_i\sim \mathcal N(0,\sigma_{\varepsilonb_i}^2 \Ib)$, and $S$ the number of pixels of an image, we have:
\beq
p(\gb_i|\fb_i,\varepsilonb_i ) 
=  \left( \frac{1}{2\pi \sigma_{\varepsilonb_i}^2} \right)^{\frac{S}{2}} \exp \left\{-\frac{1}{2\sigma_{\varepsilonb_i}^2}||\gb_i-\fb_i||^2 \right\},
\nonumber
\eeq
where $\sigma_{\varepsilonb_i}^2 \sim  \mathcal{IG}(\alpha^{\varepsilonb_i}_0,\beta^{\varepsilonb_i}_0)$ with  unknown fixed  parameters $\alpha^{\varepsilonb_i}_0$ and $\beta^{\varepsilonb_i}_0$
(inverse gamma is conjugate prior for a random variance in the Gaussian case).  

To assign $p(\fb_i|\zb,.)$ we first define the sets of pixels which are in the same class:
\begin{eqnarray*}
R_k & = & \{r : z(r) = k\}, \qquad |R_k|=n_k \\
{\fb_i}_k & = & \{ f_i(r) : z(r)=k \}.
\end{eqnarray*}
We assume that all the pixels ${\fb_i}_k$ of an image $\fb_i$ which are in the same class $k$ will be Gaussian with a random mean ${m_i}_k$  and a random variance ${\sigma_i^2}_k$, i.e.
\beq
f_i(r)|z(r)=k,{m_i}_k,{\sigma_i^2}_k \sim \mathcal N({m_i}_k,{\sigma_i^2}_k) \qquad \forall r \in \Sc ,
\nonumber
\eeq
\noindent
With these notations we have :
\beq
p({\fb_i}_k|{m_i}_k,{\sigma_i^2}_k )  =  \left( \frac{1}{\sqrt{2\pi {\sigma_i^2}_k}}\right)^{n_k} \exp \left\{ - \frac{1}{2 {\sigma_i^2}_k} || {\fb_i}_k-{m_i}_k \bm{1} ||^2 \right\},
\nonumber
\eeq
\noindent
and thus for $i=1,\dots,M$
\begin{eqnarray*}
\label{f|z}
p(\fb_i |\zb,{m_i}_k,{\sigma_i^2}_k )  
& = & \prod_{k=1}^K \left( \frac{1}{\sqrt{2\pi {\sigma_i^2}_k}}\right)^{n_k} \exp \left\{ - \frac{1}{2 {\sigma_i^2}_k} || {\fb_i}_k-{m_i}_k \bm{1} ||^2 \right\},
\end{eqnarray*}
\noindent
where $\bm{1}$ is a vector with all components equal to $1$, ${\sigma_i^2}_k\sim \mathcal{IG}({\alpha_i}_0,{\beta_i}_0)$, with  unknown fixed  parameters, and ${m_i}_k$ is an autoregressive of order 1, $AR(1)$, for each class $k$ i.e.
\begin{eqnarray}
\label{AR}
{m_i}_k & = &  \phi_k {m_{i-1}}_k + \eta_{ik},
\end{eqnarray}
where $\eta_{ik} \sim \mathcal N(0,{\sigma^2_i}_0)$, 
$\phi_k$ and ${\sigma^2_i}_0$ are unknown fixed  parameters.  Therefore
$${m_i}_k|{m_{i-1}}_k\sim \mathcal N(\phi_k {m_{i-1}}_k ,{\sigma^2_i}_0).$$

The assumption of (\ref{AR}) is the main difference of this paper with \cite{Feron04a}, i.e.
in  hyperspectral images the pixel values of a class $k$, in each channel, are not independent.

Using the relation (\ref{eq3}) 
and the density $p(\fb_i |\zb,{m_i}_k,{\sigma_i^2}_k)$ and $p(\varepsilon_i)$,
we can calculate $p(\gb_i|\zb,.)$, i.e.
\beq
\label{g|z}
\gb_i|\zb,{m_i}_k,{\sigma_i^2}_k,\sigma_{\varepsilon_i}^2 \sim \mathcal N({m_i}_k,{\sigma_i^2}_k+\sigma_{\varepsilon_i}^2).
\eeq
Finally we have to assign $P(\zb)$. 
As we introduced the hidden variable $\zb$ for finding statistically homogeneous regions in images, it is natural to define a spatial dependency on these labels. The simplest model to account for this desired local spatial dependency is a Potts Markov Random Field model:
\beq
\label{z}
P(\zb) = \frac{1}{T(\alpha)} \exp \left\{ \alpha \sum_{r\in \mathcal S} \sum_{s \in \Vc(r)}\delta (z(r) - z(s)) \right\},
\eeq
where $\mathcal S$ is the set of pixels, $\delta(0)=1$, $\delta(t)=0$ if $t\neq0$, $\mathcal V(r)$ denotes the neighborhood of the pixel $r$ (here we consider a neighborhood of 4 pixels), $T(\alpha)$ is the partition function or the normalization constant and $\alpha$ represents the degree of the spatial dependency of the variable $\zb$. 


\section{ESTIMATION USING MCMC}
\label{estimation}

Let   $\mb_i=({m_i}_k)_{k=1,\dots,K}$ and  $\sigmab_i^2=({\sigma_i^2}_k)_{k=1,\dots,K}$  be the means and the variances of the pixels in different regions of the images $\fb_i$. We define $\thetab_i$ as the set of all the parameters which must be estimated in the Bayesian framework:
\beq
\thetab_i = (\sigma_{\varepsilon_i}^2,\mb_i,\sigmab_i^2), \quad i=1,\dots,M
\nonumber
\eeq
and we note $\thetabu = (\thetab_i)_{i=1,\dots,M}$.
Now we can write the expression of the joint posterior of $p(\fbu,\zb,\thetabu|\gbu)$ by using the relations in the previous section. 
Then we propose the following Gibbs sampler,
\begin{eqnarray*}
\fbu & \sim &
\fbu|\gbu,\zb,\thetabu  \\
\zb & \sim &
\zb|\thetabu,\gbu  \\
\thetabu & \sim &
\thetabu|\fbu,\gbu,\zb
\end{eqnarray*}
\noindent
to generate  samples  $(\fbu,\zb,\thetabu)^{(1)},(\fbu,\zb,\thetabu)^{(2)}, \cdots$, and use them to compute any statistics (such as mean or median).
We may note that in each of the previous Gibbs sampling steps, we again use Gibbs scheme to sample. For example $\fbu|\gbu,\zb,\thetabu$ by alternate sampling of $\fb_i|\gb_i,\zb,\thetab_i$. This procedure is also valid for  $\thetabu$.

 It can be shown that 
$f_i(r)|g_i(r),z(r),\thetab_i$ has a Gaussian distribution and it can be sampled very easily.
On the other hand,
\beq
p(\thetab_i|\fb_i,\gb_i,\zb) \propto p({\sigma_{\varepsilon_i}^2} | \fb_i,\gb_i)\hbox{ }p(\mb_i,\sigmab^2_i | \fb_i,\zb).
\nonumber
\eeq
For the last term $p(\mb_i,\sigmab^2_i | \fb_i,\zb)$ we have to use a Gibbs algorithm and then sample following the conditional distributions $p(\mb_i|\sigmab^2_i,\fb_i,\zb)$ and $p(\sigmab^2_i |\mb_i, \fb_i,\zb)$.
It can be shown that $\sigma_{\varepsilonb_i}^2 |\fb_i,\gb_i$ and 
${\sigma_i^2}_k |\fb_i,\zb$ have inverse gamma distributions and ${m_i}_k|\fb_i,\zb,{\sigma_i^2}_k,{m_{i-1}}_k\sim \mathcal N({\mu_i}_k,{v_i^2}_k)$, with
\begin{eqnarray}
\label{postm}
{\mu_i}_k  & =  & {v_i^2}_k \left( \frac{{\phi_k {m_{i-1}}_k}}{{\sigma^2_i}_0} + \frac{1}{{\sigma^2_i}_k} \sum_{r \in R_k} f_i(r)   \right), \cr
{v_i^2}_k  & = & \left( \frac{n_k}{{\sigma_i^2}_k} + \frac{1}{{\sigma^2_i}_0}\right)^{-1}.
\end{eqnarray}
Note that if ${m_i}_k$s are independent as it is the case of multispectral images in \cite{Feron04a} then we had
$$
{\mu_i}_k   =   {v_i^2}_k \left( \frac{m_{i0}}{{\sigma^2_i}_0} + \frac{1}{{\sigma^2_i}_k} \sum_{r \in R_k} f_i(r)   \right),$$
where ${m_{i}}_0$ is a fixed number. 

Finally, we can write the posterior 
probability of $\zb$ by
\beq
\label{postz}
P(\zb | \gbu,\thetabu)  \propto  p(\gbu |\zb,\thetabu)~P(\zb) 
 =  \prod_{i=1}^M p(\gb_i|\zb,\thetab_i) ~P(\zb),
\eeq
where can be calculated by using (\ref{g|z}) and (\ref{z}). 
As we choose a Potts Markov Random Field model for the labels $\zb$, we may note that an exact sampling of the \emph{a posteriori} distribution $P(\zb | \gbu,\thetabu)$ is  impossible. But we can  still use a Gibbs sampling to generate parallel samples of $\zb$.

For simplicity sake, we estimate the parameters $\phi_k$, ${k=1,\dots,K}$ 
 with a classical method and 
we consider it as constant in this section. If the series 
$\{{m_i}_k\}_{i=1,\dots,M}$  has an $AR$ model, ($k$ is fixed), then we can estimate $\phi_k$ efficiently, because the number of images $M$ is large. 
\par
Here we give the summary of the proposed algorithm for estimating $\zb$ which has the following steps:
\begin{enumerate}
\item  Find an initial joint segmentation of the hyperspectral image by using
any simple segmentation or classification method such as k-means method,
\item Calculate $\{{m_i}_k\}_{i=1,\dots,M}$ for ${k=1,\dots,K}$,
\item Fit an AR model for each  series 
$\{{m_i}_k\}_{i=1,\dots,M}$, ${k=1,\dots,K}$ with a classical method,
\item Use the proposed  Gibbs algorithm to generate samples of $\{{m_i}_k\}$ using (\ref{postm}) and $\zb$ using  (\ref{postz}).
\end{enumerate}

\section{Simulation }
\label{results}

\subsection{Synthetic data}
To measure the performance of the proposed method, first we generate artificially a set of data, starting by a known segmentation   $\zb$ 
and  generate the images $\fb_i |\zb=k$ as  homogeneous Gaussian with known mean  ${m_i}_k$ and variance ${\sigma_i^2}_k $ with 
${m_i}_k  =   \phi_k {m_{i-1}}_k + \eta_{ik}$, where $|\phi_k|<1$  is a fixed value. Then we generate the  $\gb_i$ by adding a Gaussian noise.
Finally we use these data as input for different segmentation methods and compare their relative performance. 

In this example, we know the original $\zb$. Therefore, comparison can be done by counting the number of misclassified pixels. 

Figure 1 shows the results of this experiment.
The first row of Figure~1 shows $\zb$,  $\fbu$, and $ \gbu$.
The second row shows the results of  segmentation of images $ \gbu$
by two methods: the first assume independence of ${m_i}_k$ ($i=1,\cdots, M)$ and
second which is the proposed method use an AR modeling of ${m_i}_k$. The number of miss classified pixels for $\hat \zb_1$ is more than 1000 and for 
$\hat \zb_2$ is less than 200.

\newpage
\begin{figure}[!htb]
\btabu{c}
\includegraphics[width=55mm,height=55mm]{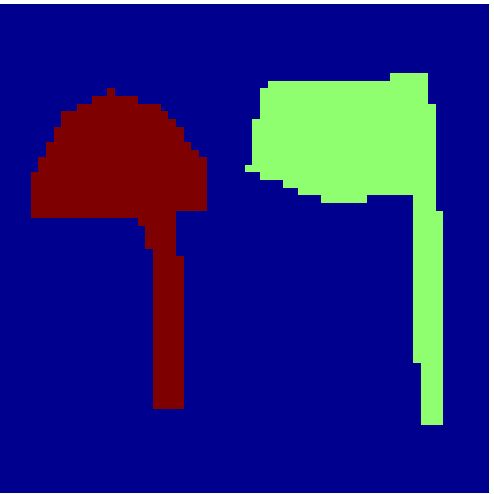} 
\\ $\zb$ \\ 
\btabu{@{}c@{}c@{}}
\includegraphics[width=70mm,height=70mm]{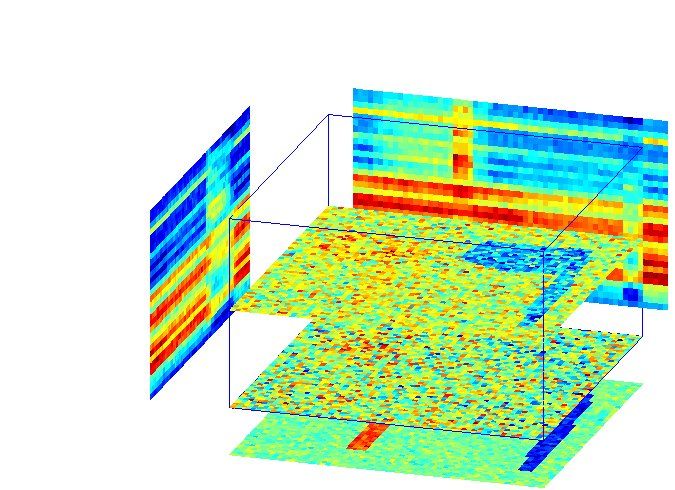} &
\includegraphics[width=70mm,height=70mm]{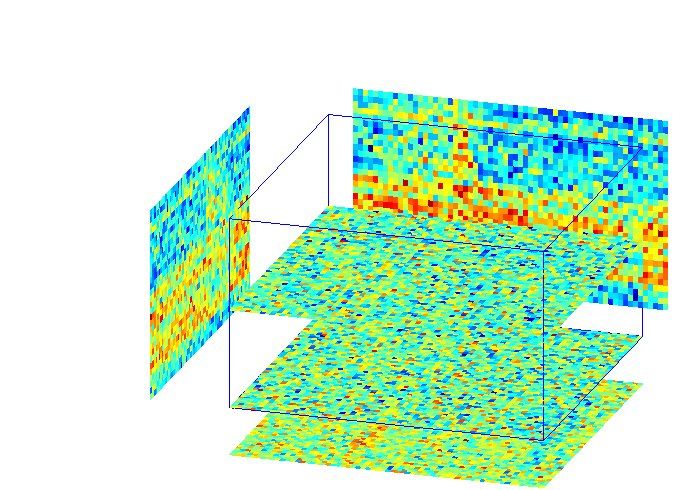}		
\\
$\fbu$ and their marginals  & 
$\gbu$ and their marginals
\\
\includegraphics[width=55mm,height=55mm]{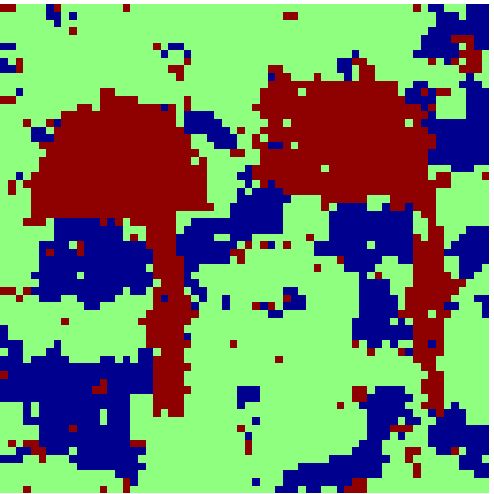} &
\includegraphics[width=55mm,height=55mm]{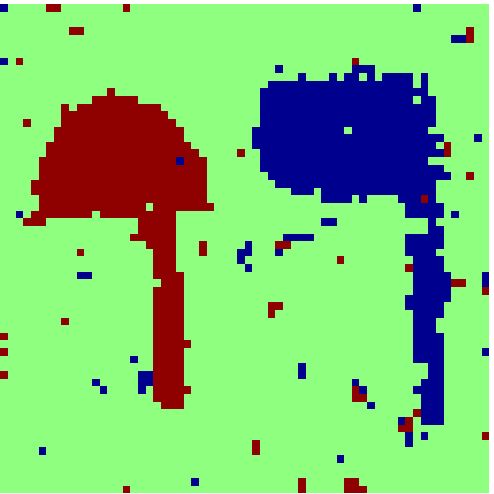}
\\
$\hat \zb_1$ with independence and &
$\hat \zb_2$,	with AR assumption on  $\{{m_i}_k\}$
\etabu
\etabu
\caption{Bayesian segmentation of a simulated hyperspectral image with independence and 	 AR assumption on  $\{{m_i}_k\}$ with 50 iterations.   }
\end{figure}

\newpage
\subsection{Real data}
In the next step we applied the proposed method to a part of real data (Aviris).
In Figure~2
 we illustrate a real example hyperspectral images $\gbu$ which are 56, ($128\times 124$) images.  This figure  shows the reconstruction and joint segmentation results  i.e. $\hat \zb_1$ with {\em independence} and  
$\hat \zb_2$,	with {\em AR} assumption on  $\{{m_i}_k\}$ after 200 iterations.
Unfortunately, in this case we cannot give a quantitative comparing of the results.

\begin{figure}[!htb]
\begin{tabular}{c}
\includegraphics[width=85mm,height=85mm]{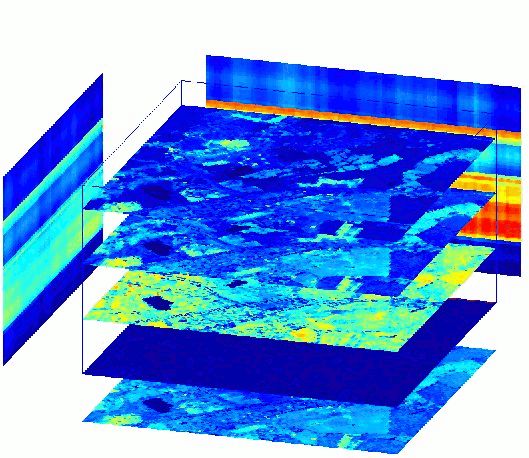} \\ 
$\gbu$ \mbox{ and their marginals} \\ ~\\ 
\begin{tabular}{cc}
		\includegraphics[width=55mm,height=55mm]{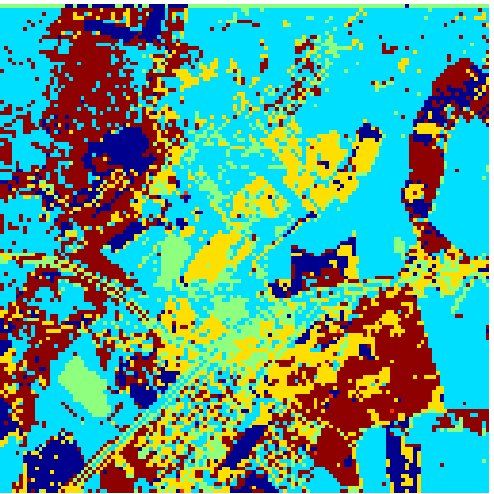} &
		\includegraphics[width=55mm,height=55mm]{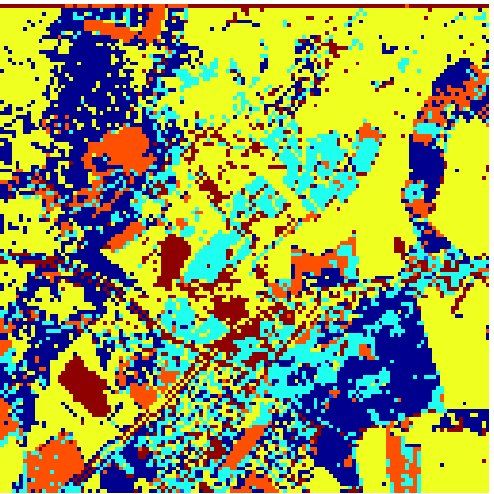}		
	\\  
	$\hat \zb_1$ with {\em independence} and  
	&$\hat \zb_2$,	with {\em AR} assumption on  $\{{m_i}_k\}$ \\ ~\\
	\end{tabular} 
\end{tabular} 
\caption{Bayesian segmentation of a part of real hyperspectral image with dimension $128\times 128 \times  56$ with
independence and 	 AR assumption on  $\{{m_i}_k\}$ with 200 iterations.
 }	
\end{figure}

\newpage

\end{document}